\documentstyle[twoside,fleqn,espcrc2,epsf]{article}

\newcommand{\AmS}{{\protect\the\textfont2
  A\kern-.1667em\lower.5ex\hbox{M}\kern-.125emS}}

\newcommand{\bd}{\begin{displaymath}}
\newcommand{\ed}{\end{displaymath}}
\newcommand{\be}{\begin{equation}}
\newcommand{\ee}{\end{equation}}
\newcommand{\bea}{\begin{eqnarray}} 
\newcommand{\eea}{\end{eqnarray}}
\newcommand{\bt}{\begin{tabular}}
\newcommand{\et}{\end{tabular}\newline}

\title{
\vspace{-3.15cm}                                            
{\normalsize MIT-CTP-3170}    \\[-0.2cm]                    
{\normalsize August 2001} \\                                
\vspace{2.25cm}                                             
Perturbative renormalization for overlap fermions}

\author{
Stefano Capitani\address{Center for Theoretical Physics, 
Laboratory for Nuclear Science, \\
Massachusetts Institute of Technology,
77 Massachusetts Ave., Cambridge, MA 02139, USA}
}

\begin{document}

\begin{abstract}
Using lattice overlap fermions, we have computed the 1-loop renormalization 
factors of several operators that measure DIS structure functions and weak
amplitudes. Computer codes written in the algebraic manipulation language 
FORM have been used. The improvement of the operators is also discussed.
\end{abstract}

\maketitle

\section{INTRODUCTION}

Overlap fermions~\cite{overlap1,overlap2} exhibit an exact chiral symmetry 
on the lattice also for non-zero lattice spacing~\cite{symmetry}. 
Recent numerical works show the quite good accuracy with which chiral 
symmetry is attained in the overlap formulation~\cite{numerical1,numerical2}.

The overlap-Dirac operator that we use is 
\be
D_N = \frac{\rho}{a} \, \Big[ 1+ \frac{X}{|X|} \Big] 
,~X=D_W -\frac{\rho}{a} ,~0 <\rho <2,
\ee
where $D_W$ is the Wilson-Dirac operator ($r$ =1).

\section{RENORMALIZATION}

Monte Carlo computations of matrix elements of operators must be 
renormalized in order to obtain physical results. Lattice simulations 
need thus to be supported by the knowledge of the renormalization 
factors of the relevant operators.

Operator mixing for overlap fermions is simpler than for Wilson fermions. 
Chiral symmetry prohibits any mixings among operators of different chirality
and in general reduces the number of operators which mix. Operators whose
mixing coefficients are power-divergent like $a^{-n}$ in the continuum 
limit do not mix in the overlap if they belong to multiplets with the 
wrong chirality.

The renormalization of the bilinears $\bar{\psi} \Gamma \psi$ with overlap 
fermions was first computed by Alexandrou et al.~\cite{afpv}. After that
we have calculated the renormalization of several operators measuring the 
lowest moments of DIS structure functions~\cite{sf}.
We have considered the unpolarized quark distribution $q$, the helicity 
distribution $\Delta q$, the transversity distribution $\delta q$ and 
also the $g_2$ structure function. Their moments are proportional to 
the hadronic matrix elements of the towers of operators
\bea
\langle x^n \rangle_q   & \sim &  \langle \vec{p}, \vec{s} | 
\bar{\psi} \gamma_{\{\mu} D_{\mu_1} \cdots D_{\mu_n\}} \psi 
| \vec{p}, \vec{s} \rangle   \\
\langle x^n \rangle_{\Delta q}  & \sim &  \langle \vec{p}, \vec{s} |  
\bar{\psi} \gamma_{\{\mu} \gamma_5 D_{\mu_1} \cdots D_{\mu_n\}} \psi 
| \vec{p}, \vec{s} \rangle   \nonumber \\
\langle x^n \rangle_{g_2}  & \sim &  \langle \vec{p}, \vec{s} |
\bar{\psi} \gamma_{[\mu} \gamma_5 D_{\{\mu_1} D_{\mu_2 ]}
\cdots D_{\mu_n\}} \psi | \vec{p}, \vec{s} \rangle    \nonumber \\
\langle x^n \rangle_{\delta q}  & \sim &  \langle \vec{p}, \vec{s} |  
\bar{\psi} \sigma_{\mu \{ \nu} \gamma_5 D_{\mu_1} \cdots D_{\mu_n\}} \psi 
| \vec{p}, \vec{s} \rangle  . \nonumber
\eea
In Table 1 we show the $Z$'s of the multiplicatively renormalized 
operators for two values of the parameter $\rho$ (and also for Wilson), 
for $\beta=6.0$, in the $\overline{\rm{MS}}$ scheme. In some cases for 
a given moment we have computed two operators ($(a)$ and $(b)$), 
which belong to two different representations of the discrete Euclidean 
Lorentz group. In the operators $(a)$ all Lorentz indices are distinct.

In the Wilson case the operators measuring the moments of the $g_2$ 
structure function are not multiplicatively renormalized, and the 
additional mixings with wrong-chirality operators have power-divergent 
coefficients. Moreover, the $Z$'s of the corresponding moments of 
$q$ and $\Delta q$ are not constrained to be equal anymore.

It can be seen in Table 1 that the renormalization factors for $\rho=1.0$ 
are large. Contrary to Wilson fermions, most of the renormalization comes 
from the quark self-energy, while the contributions of the remaining 
diagrams are small, close to the corresponding Wilson results and 
depend very little on $\rho$. The quark self-energy instead decreases 
as $\rho$ increases, and this suggests to consider higher $\rho$'s 
to get smaller $Z$'s. The $Z$'s for $\rho=1.9$ are indeed smaller 
than for $\rho=1.0$, and closer to the Wilson results (see Table 1).
\begin{table}
\caption{Renormalization constants of operators measuring moments of 
         structure functions, for $\beta=6.0$.}
\begin{tabular}{|c|r|r|r|} \hline
moment   &  \multicolumn{2}{|c|}{overlap}      &  Wilson   \\ 
         &  \multicolumn{1}{|r}{$\rho=1.0$}    
         &  \multicolumn{1}{r|}{$\rho=1.9$}    &           \\ \hline
$~~\langle x \rangle_q^{(a)}~~$ 
&~~1.41213  &~~1.21841  &~~0.98920  \\
$ \langle x \rangle_{\Delta q}^{(a)} $ &
1.41213  &  1.21841  &  0.99709  \\
$ \langle x \rangle_q^{(b)} $ &
1.40847  &  1.21309  &  0.97837  \\
$ \langle x \rangle_{\Delta q}^{(b)} $ &
1.40847  &  1.21309  &  0.99859  \\
$ \langle x^2 \rangle_q $ &
1.51968  &  1.32436  &  1.09763  \\
$ \langle x^2 \rangle_{\Delta q} $ &
1.51968  &  1.32436  &  1.10231  \\
$ \langle x^3 \rangle_q^{(a)} $ &
1.61872  &  1.42279  &  1.19722  \\
$ \langle x^3 \rangle_{\Delta q}^{(a)} $ &
1.61872  &  1.42279  &  1.20040  \\
$ \langle x^3 \rangle_q^{(b)} $ &
1.63737  &  1.44159  &  1.21534  \\
$ \langle x^3 \rangle_{\Delta q}^{(b)} $ &
1.63737  &  1.44159  &  1.21944  \\
$ \langle x \rangle_{g_2} $ &
1.34794  &  1.18456  &  mixing      \\
$ \langle x^2 \rangle_{g_2} $ &
1.47816  &  1.30997  &  mixing      \\
$ \langle x^3 \rangle_{g_2} $ &
1.58943  &  1.41900  &  mixing      \\
$ \langle 1 \rangle_{\delta q} $ &
1.27252  &  1.08648  &  0.85631  \\
$ \langle x \rangle_{\delta q} $ &
1.41153  &  1.21851  &  0.99559  \\
$ \langle x^2 \rangle_{\delta q} $ &
1.51865  &  1.32355  &  1.10021  \\
\hline
\end{tabular} 
\end{table}

The renormalization of the four-fermion operators of the $\Delta F=2$ 
and $\Delta S=1$ effective weak Hamiltonians has been studied together 
with L.~Giusti~\cite{weak}. They describe physics like the 
$K^0$-$\bar K^0$ and $B^0$-$\bar B^0$ mixings, the $\Delta I=1/2$ rule 
(octet enhancement) and the $CP$ violation parameter $\epsilon'/\epsilon$. 
The two latter cases are not easy to study with Wilson fermions, 
because of power-divergent operators that occur under renormalization
and which have to be non-perturbatively subtracted. Neuberger's fermions 
are appealing because the exact chiral symmetry forbids many mixings 
which occur in the Wilson case. The GIM mechanism, which as a 
consequence of chiral symmetry is quadratic instead of linear 
in the masses, is as powerful as in the continuum in eliminating 
unwelcome operators. Furthermore, corresponding parity-conserving 
and parity-violating operators are renormalized in the same way.

With overlap fermions the renormalization of the operators relevant 
for the $\Delta I=1/2$ rule can be done without any power-divergent 
subtractions. The calculation of $\epsilon'/\epsilon$ is also greatly 
simplified compared to the Wilson case, although one power-divergent 
mixing still remains. We only outline here the renormalization of the 
$\Delta I=1/2$ amplitudes, described by the operators
\be
{\cal O}_\pm = (Q_1 -Q^c_1) \pm (Q_2 -Q^c_2),
\ee
where~\footnote{Here $\gamma^\mu_{L,R} = \gamma^\mu(1\mp\gamma_5)$ 
and $\alpha,\beta$ are color indices.}
\bea
Q_1    =  \bar{s}^\alpha \gamma^\mu_L u^\beta \cdot
          \bar{u}^\beta  \gamma^\mu_L d^\alpha  &,& \!\!
Q_2    =  \bar{s} \gamma^\mu_L u \cdot
          \bar{u} \gamma^\mu_L d               \\
Q^c_1  =  \bar{s}^\alpha \gamma^\mu_L c^\beta \cdot
          \bar{c}^\beta  \gamma^\mu_L d^\alpha  &,& \!\!
Q^c_2  =  \bar{s} \gamma^\mu_L c \cdot
          \bar{c} \gamma^\mu_L d .             \nonumber 
\eea
The flavor structure forbids mixings between ${\cal O}_{+}$ and 
${\cal O}_{-}$ and chiral symmetry forbids mixings with other 
dimension-six operators. The complete renormalization is given by
\bea
\widehat O_\pm (\mu) & = & Z_\pm (\mu a, g^2_0) 
\widetilde {\cal O}_\pm (a) + O(a^2) \label{eq:opm} \\
\widetilde {\cal O}_\pm (a) & = & {\cal O}_\pm (a) + (m^2_c -m^2_u) 
C_\pm^m(g_0^2) {\cal Q}_m(a) \nonumber ,
\eea
where ${\cal Q}_m = (m_s+m_d) \bar{s} d + (m_s-m_d) \bar{s} \gamma_5 d$.
Thanks to the quadratic GIM factor $m_c^2-m_u^2$, the mixing coefficients 
$C_{\pm}^m$ are finite. In principle the operators 
${\cal O}_\pm$ would mix also with 
${\cal Q}_\sigma = g_0 [(m_s+m_d)\bar{s} \sigma_{\mu\nu} F_{\mu\nu} d
+ (m_s-m_d) \bar{s} \sigma_{\mu\nu} \widetilde F_{\mu\nu} d]$, 
however thanks to the quadratic GIM mechanism and to the 
$m_s \pm m_d$ factors coming from chiral symmetry these mixing 
coefficients are of $O(a^2)$.

In a regularization that breaks chiral symmetry, like Wilson, the GIM 
mechanism is only linear in the masses. Furthermore, the parity-conserving 
and parity-violating components of ${\cal Q}_m$ and ${\cal Q}_\sigma$ 
behave differently; in particular, the $m_s+m_d$ factors are absent. 
The parity-conserving part of $C_{\pm}^m$ is then quadratically 
divergent~\footnote{The $m_s-m_d$ factors are present also for Wilson fermions 
thanks to $CPS$ symmetry ($S$ is the interchange ($s\leftrightarrow d$)).}.

The $Z_{\pm}$ factors in (\ref{eq:opm}) are simple linear combinations of 
$Z_S$, $Z_V$ and $Z_\psi$~\cite{weak}. The $C_\pm^m$ coefficients are 
not needed for the physical $K\rightarrow \pi\pi$ matrix elements; if 
$K\rightarrow \pi$ amplitudes are used, they can be determined by a 2-loop 
calculation or non-perturbatively using $K\rightarrow 0$ matrix elements.

All calculations have been done using codes written in the algebraic 
manipulation language FORM. The gauge invariance of the $Z$'s and the 
implementation of dimensional regularization (NDR and 't~Hooft-Veltman) 
and a mass regularization allow strong checks of the calculations, which
we also did in some cases by hand. We also checked the Wilson results 
when known, and when not we computed them for the first time~\cite{sf}.

\section{IMPROVEMENT}

Overlap fermions present many advantages compared to Wilson fermions
also when the issue is the Symanzik improvement of operators. 

Although Neuberger's action (and therefore the spectrum of the theory) 
is already $O(a)$ improved, matrix elements of operators have in general 
$O(a)$ corrections. Operators of the form $O = \bar{\psi} \widetilde{O} \psi$
are improved by considering~\cite{qcdsf1,qcdsf2} 
\be
O^{\rm imp} = \bar{\psi} \Big(1-\frac{1}{2\rho} a D_N\Big) \, \widetilde{O}
\, \Big(1-\frac{1}{2\rho} a D_N\Big) \psi ,
\label{eq:opimp}
\ee
which works to all orders of perturbation theory. 

Compared with Wilson fermions, the $O(a)$ improvement for Neuberger's 
fermions presents a few very convenient simplifications:
\begin{enumerate}
\item the action is already improved, and there is no need of 
      new interactions like for Wilson fermions (Sheikholeslami-Wohlert);
\item the renormalization constants for improved and unimproved operators 
      are the same~\cite{afpv}, while with Wilson fermions additional 
      cumbersome calculations are needed to get the $Z$'s 
      for the improved operators; 
\item the improved operators are always given by Eq.~(\ref{eq:opimp}), 
      while for Wilson fermions the construction of the improved operator 
      is different in each case;
\item full $O(a)$ improvement is achieved without tuning any coefficients,
      and it is valid to all orders of perturbation theory.
\end{enumerate} 
The last point is really a big calculational advantage. In fact, improving
an operator with Wilson fermions means first finding out a complete basis 
of operator counterterms, and then determining for all counterterms the 
exact values of their coefficients that effectively improve the original 
operator, order by order in perturbation theory. 
This appears even at lowest order to be a highly demanding task,
as it can be seen for the first moment of unpolarized structure functions, 
$\bar{\psi} \gamma_{\{\mu} D_{\nu\}} \psi$. 
In this case two counterterms are needed,
\bea
O^{(1)}_{\mu \nu} &=& -\frac{1}{4} a {\rm i} c_1 (g_0^2) \, \sum_\lambda
  \bar{\psi}\sigma_{\lambda\{\mu} \Big[ D_{\nu\}} , D_\lambda \Big] \psi 
\nonumber \\
O^{(2)}_{\mu \nu} &=& -\frac{1}{4} a c_2 (g_0^2) \, \bar{\psi} 
  \Big\{ D_\mu , D_\nu  \Big\} \psi , 
\eea
and up to now it has been possible to determine at 1 loop only one of the 
corresponding coefficients, while the other one remains unknown~\cite{qcdsf2}.

\section{CONCLUSIONS}

The overlap renormalization factors of most operators necessary in the study
of DIS and weak processes are now known. The exact chiral symmetry of overlap
fermions makes the study of long-standing problems like the $\Delta I=1/2$ 
amplitudes and $\epsilon'/\epsilon$ much more accessible than before. 

This work has been supported in part by the U.S. Department of Energy (DOE) 
under cooperative research agreement DE-FC02-94ER40818. I thank L.~Giusti
for the pleasant collaboration.

\end{document}